%% file: CP_2025_IZAWOL.tex
\pdfoutput=1
\documentclass[11pt]{article} 
\usepackage{amsfonts,amssymb,amsmath,amsthm,dsfont,graphicx,hyperref,mathrsfs,euscript,enumitem,framed,bm, soul, bbm, booktabs, array,multicol, multirow}
\usepackage[title, titletoc]{appendix}
\usepackage[onehalfspacing]{setspace}
\usepackage{ifthen,caption,subfigure,lscape}
\hypersetup{pdfborder = {0 0 0},colorlinks=true,linkcolor=blue,urlcolor=blue,citecolor=blue}
\usepackage[numbers]{natbib}

\usepackage[top=1in,bottom=1in,left=1in,right=1in]{geometry}

\numberwithin{equation}{section}

\allowdisplaybreaks[1]

\renewcommand*{\arraystretch}{.6}

\begin{document}

{\singlespacing
\title{\vspace{-0.0in}Leveraging Covariates in Regression Discontinuity Designs\footnote{Invited contribution prepared for the \textit{\href{https://wol.iza.org/}{IZA World of Labor}}. We thank two anonymous reviewers and the editor, Arnaud Chevalier, for their comments.}\medskip}
\author{Matias D. Cattaneo\thanks{Department of Operations Research and Financial Engineering, Princeton University.}
        \and
	    Filippo Palomba\thanks{Department of Economics, Princeton University.}
        }
\maketitle
}

\pagestyle{plain}


\section*{Keywords}

Causal inference, treatment effect estimation, regression discontinuity, covariate adjustment, head start.

\section*{Teaser}

Proper use of covariates in Regression Discontinuity designs can enhance empirical scientific discoveries and evidence-based policy decisions.

\section*{Elevator Pitch}

It is common practice to incorporate additional covariates in empirical economics. In the context of Regression Discontinuity (RD) designs, covariate adjustment plays multiple roles, making it essential to understand its impact on analysis and conclusions. Typically implemented via local least squares regressions, covariate adjustment can serve three main distinct purposes: (i) improving the efficiency of RD average causal effect estimators, (ii) learning about heterogeneous RD policy effects, and (iii) changing the RD parameter of interest.

\clearpage
\section*{Graphical Abstract}

\begin{figure}[!ht]
    \centering
    \includegraphics[width=0.9\linewidth]{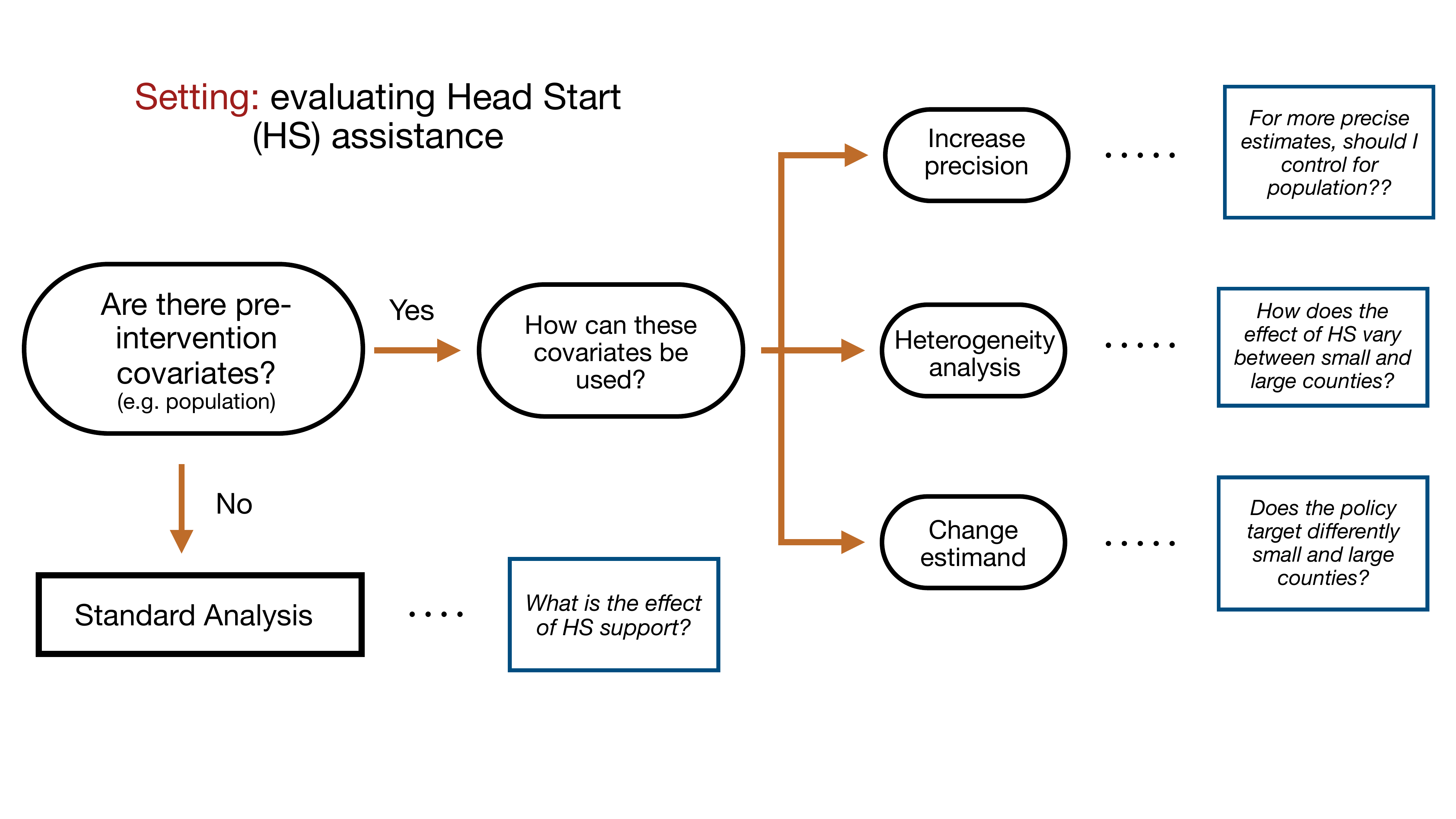}
\end{figure}

\section*{Key Findings}

\begin{itemize}
    \item \textbf{Pros}
    \begin{itemize}[leftmargin=*, label=$+$]
        \item Precision of RD causal effect estimators can be improved by correctly using pre-intervention covariates.
        \item When properly implemented, covariate adjustment can uncover interesting heterogeneous RD causal effects.
        \item Various other RD policy effects of interest can be learned by correctly incorporating covariates in the analysis.
    \end{itemize}
    
    \item \textbf{Cons}
    \begin{itemize}[leftmargin=*, label=$-$]
        \item Understanding empirical results based on covariate-adjusted RD estimates requires careful consideration.
        \item Covariate adjustment cannot restore the validity of an RD design without strong parametric assumptions.
        \item Empirical work leveraging covariates in RD designs is often undisciplined and ad-hoc, potentially leading to invalid empirical findings and policy prescriptions. 
    \end{itemize}
\end{itemize}

\section*{Authors' main message}

When applied correctly, covariate adjustment in RD designs can significantly enhance empirical analysis and strengthen policy conclusions. However, the use of covariates across RD studies is often inconsistent and ad hoc, undermining both the credibility and replicability of findings. Adopting best methodological practices for covariate adjustment in RD designs can improve efficiency and support rigorous heterogeneity analysis. Additionally, pre-intervention covariates can be leveraged to modify the RD parameter of interest, though this requires imposing additional, sometimes stringent, modeling assumptions.


\section*{Motivation}

In program evaluation, a primary goal is to determine the causal effect of a policy on an outcome. Among the numerous experimental and non-experimental research designs available in the literature (see \cite{Abadie-Cattaneo_2018_ARE} and references therein), the RD design is widely used in empirical economics. A landmark example of its application is found in \cite{Ludwig-Miller_2007_QJE}, which examines the causal impact of the U.S. federal anti-poverty program Head Start on child mortality. They found a positive and statistically significant overall average RD treatment effect. The Head Start data also included pre-intervention covariates that can be leveraged to improve the precision of estimating the average causal effect at the cutoff, as well as to explore treatment effect heterogeneity. In this article, we discuss the role of covariate adjustment in RD designs and demonstrate their application in the context of the Head Start study.

\section*{Background Information}

The Head Start application serves as a prototypical empirical example of the canonical (sharp) RD design. The original study leveraged a discontinuity in program funding that occurred in $1965$, when Head Start was first implemented. To ensure adequate representation of the nation's poorest communities in a nationwide grant competition for program funds, the federal government provided assistance to the $300$ poorest U.S. counties to help them prepare and submit funding applications. This support resulted in a surge of participation and funding rates among these counties, creating a discontinuity at the $300$th poorest county, and thus enabled the use of an RD design for policy evaluation. In the Head Start context, the units are counties, the score is the poverty index, and the cutoff is the poverty level of the $300$th poorest U.S. county (equal to $59.1984$).

In the sharp RD design, the strategy to identify the average treatment effect at the cutoff hinges on the following two key conditions holding at the cutoff:
\begin{enumerate}[label=(\textit{\roman*})]
    \item a jump in the treatment assignment rule; and
    \item continuity of the conditional expectation of the potential outcomes given the score.
\end{enumerate}
The first condition ensures that the treatment assignment is as if randomly assigned near the cutoff, while the second condition ensures that, had their treatment status remained unchanged, these units would exhibit the same average response. In addition, from a practical perspective, a positive density of observations near the cutoff is needed for estimation and inference. Under these conditions, comparing units just above and below the threshold leads to valid causal treatment effect estimates. Local least squares linear regression employing only those units whose scores are close to the cutoff determining policy assignment is often used to estimate the RD average causal effect at the cutoff (see \cite{Cattaneo-Idrobo-Titiunik_2020_CUP} and \cite{Cattaneo-Idrobo-Titiunik_2024_CUP} for a comprehensive practical introduction).

Using flexible parametric methods, \cite{Ludwig-Miller_2007_QJE} found that access to increased Head Start funding led to a significant reduction in child mortality rates between $1973$ and $1983$ among children aged five to nine, specifically for causes impacted by the program's health services. The reported decrease in mortality was substantial, from approximately $3.2$ to $1.9$ deaths per $100,000$ children. More recently, \cite{Cattaneo-Titiunik-VazquezBare_2017_JPAM} re-examined the Head Start data using modern RD methods (see \cite{Cattaneo-Titiunik_2022_ARE} and references therein), and showed that the reduction in child mortality is robust to different assumptions and estimation strategies.

More recent work on the Head Start data also incorporated pre-treatment covariates. Following \cite{Calonico-Cattaneo-Farrell-Titiunik_2019_RESTAT}, we leverage nine pre-intervention covariates: \textit{population}, \textit{population between age 14 and 17}, \textit{population between age 5 and 34}, \textit{population with age 25+}, \textit{\% attending school between age 14 and 17}, \textit{\% attending school between age 5 and 34}, \textit{\% of urban areas}, \textit{\% of urban areas}, and \textit{\% of Black population}. The eight covariates, excluding \textit{population}, are only used for improving estimation efficiency, and thus we refer to them as ``covariates for efficiency''. The pre-intervention covariate \textit{population} will be used for either improving estimation efficiency or heterogeneity analysis. This application cannot be used to demonstrate empirically the other possible roles of covariate adjustment in RD designs, which we also briefly discuss below. Finally, we offer practical guidance and briefly address other empirical and methodological uses of covariates in RD-based research: see \cite{Cattaneo-Keele-Titiunik_2023_HandbookCh} for a recent methodological overview and further references. General-purpose software \citep{Calonico-Cattaneo-Farrell-Palomba-Titiunik_2025_Stata} and replication files are available at \url{https://rdpackages.github.io/}.

\section*{Discussion of pros and cons}

The RD design has gained recognition as a leading research design for observational studies in the last twenty years: it was first formalized by \cite{Hahn-Todd-vanderKlaauw_2001_ECMA} from a continuity-based (local regression) perspective, and later by \cite{Cattaneo-Frandsen-Titiunik_2015_JCI} from a local-randomization (experimental) perspective. In empirical economics, the former conceptual perspective is often adopted because it justifies a local least squares regression analysis. While a wealth of methodological approaches are already available for the analysis and interpretation of RD designs, the role of covariates in particular remains an active area of methodological research. There is still no consensus on how to leverage covariates in RD applications among practitioners, a fact that has led to different valid (and invalid) empirical approaches, which can hamper the transparency and replicability of the empirical findings and policy recommendations.

Employing the Head Start application, we discuss the two main distinct roles of covariate adjustment in RD designs: for improving efficiency and for heterogeneity analysis. We also provide a unified framework to discipline practice, leveraging covariates in empirical work. Moreover, we review other roles that covariates can take in RD applications conceptually, and caution against invalid practices sometimes found in the empirical literature. 

\subsection*{Leveraging Covariates for Efficiency}

Using the continuity-based framework, \cite{Calonico-Cattaneo-Farrell-Titiunik_2019_RESTAT} first studied and formalized the role of covariate adjustment for efficiency purposes in RD designs. Their main goal was to provide an easy-to-interpret and easy-to-implement methodology to incorporate covariates in the otherwise canonical local linear regression approach for estimation of the (sharp) RD treatment effect. Building on the standard analysis of experiments, the goal was not to change the parameter of interest but rather to improve the statistical precision in estimating the canonical RD average treatment effect at the cutoff. Indeed, no additional identification assumptions are needed in this case, beyond assuming that the additional covariates are pre-intervention.

\begin{figure}[!ht]
    \caption{Head Start Canonical RD Average Treatment Effect\label{fig:RDplot-conventional}}
    \centering
    \subfigure[Global RD Plot]{
        \includegraphics[width=0.45\textwidth]{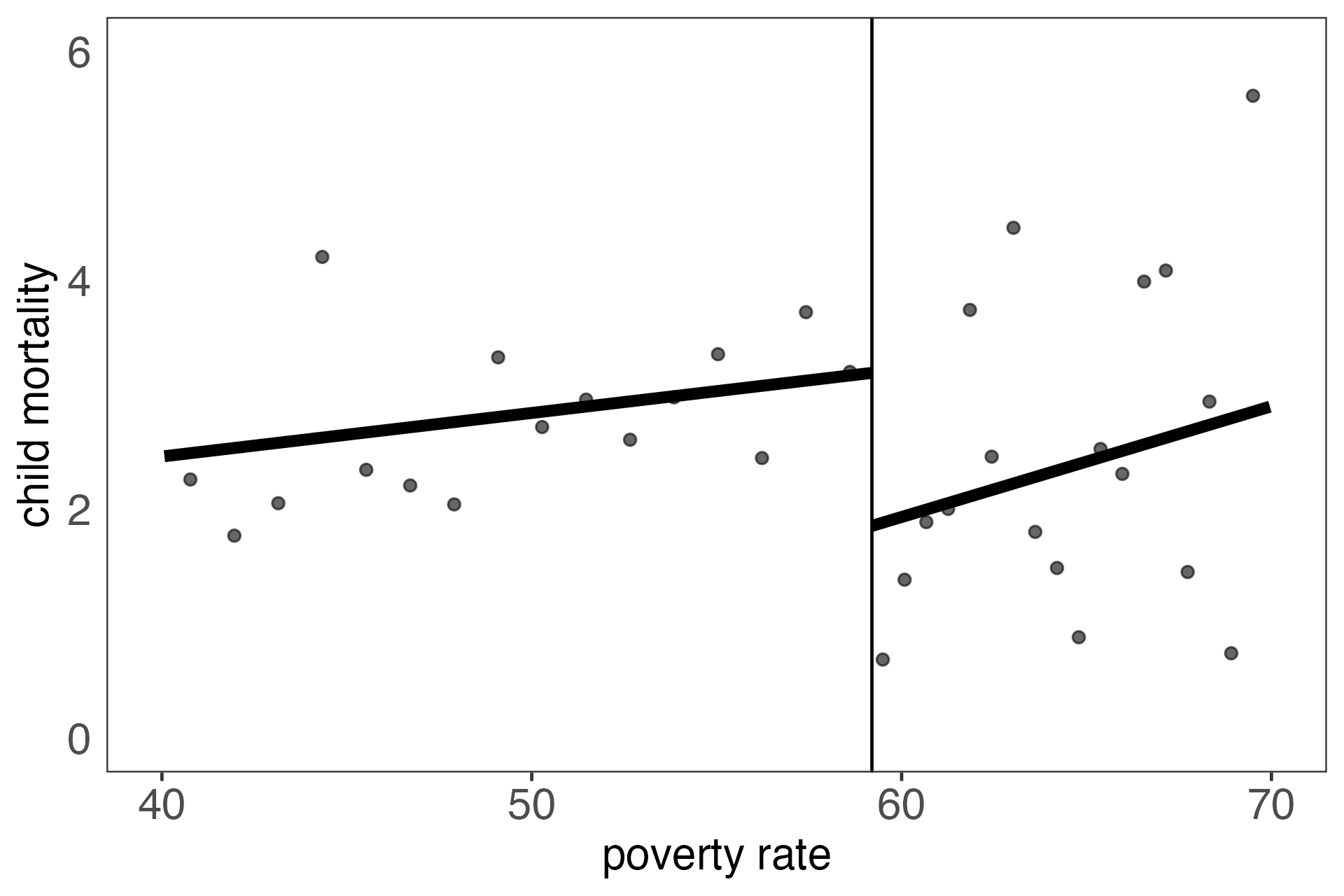}
    }
    \subfigure[Local RD Plot]{
        \includegraphics[width=0.45\textwidth]{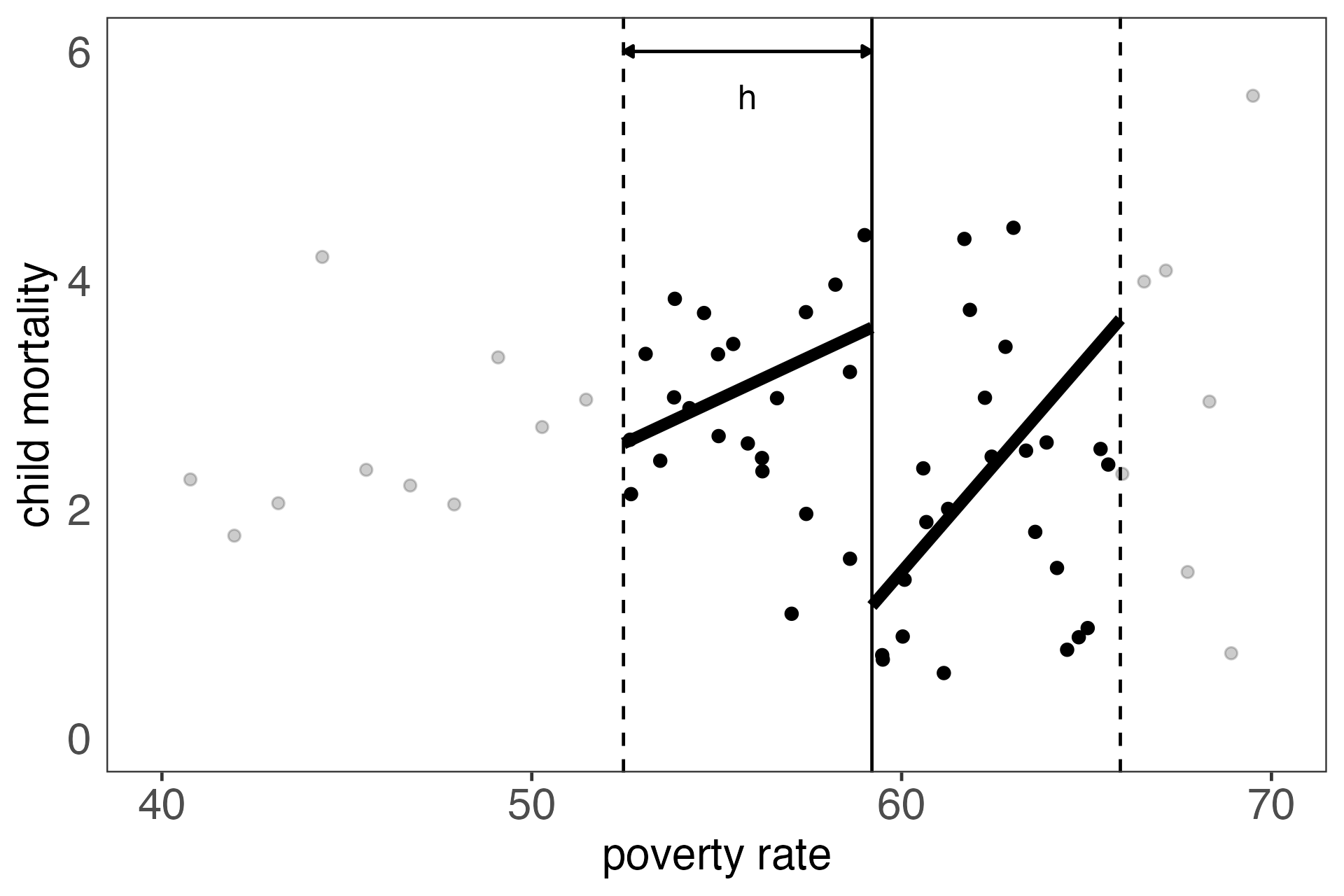}
    } 
    \par
    \begin{center}
        \parbox[1]{\textwidth}{\footnotesize \textit{Notes:} In all cases, we use a triangular kernel, employ a polynomial of degree 1 in the running variable, and equate the main bandwidth and the auxiliary bias bandwidth. No covariates are included for efficiency purposes. Dots represent averages within equally-spaced bins, solid lines represent separate linear fits on each side of the cutoff, and vertical dashed lines (when displayed) delimit the estimation sample of length twice the optimal MSE bandwidth. Sample sizes in panel (a) are $n_-=2504$ and $n_+=300$, whereas in panel (b) $N_-=231$ and $N_+=181$}.
    \end{center} 
\end{figure}

We illustrate the main idea using the Head Start application. Figure \ref{fig:RDplot-conventional}(a) presents an RD Plot employing a global linear regression, which suggests a sizable RD treatment effect. The first column in Table \ref{tab:cov-efficiency}, labeled ``canonical'', presents a modern implementation of the RD design originally studied in \cite{Ludwig-Miller_2007_QJE}. The estimated policy effect is a reduction of $2.41$pp in child mortality among those U.S. counties that were offered Head Start, which is statistically significant at the $95\%$ conventional level when using robust bias-corrected statistical inference. The estimated RD treatment effect is highly statistically significant and represents a roughly $80\%$ reduction in child mortality relative to the average mortality rate among control counties. This classical empirical finding is illustrated using a local RD Plot in Figure \ref{fig:RDplot-conventional}(b).

A key insight from \cite{Calonico-Cattaneo-Farrell-Titiunik_2019_RESTAT} is that, to increase estimation efficiency, pre-intervention covariates should be incorporated with a restricted, common parameter across both control and treatment groups to obtain a consistent and more precise estimator of the canonical (sharp) RD treatment effect, under minimal identification assumptions and with low misspecification bias. 

The last two columns in Table \ref{tab:cov-efficiency} correspond to the approach proposed by \cite{Calonico-Cattaneo-Farrell-Titiunik_2019_RESTAT} when either the eight covariates for efficiency are included in the estimation, or all nine pre-intervention covariates are incorporated in the analysis. The second column in Table \ref{tab:cov-efficiency}, labeled ``with covariates for efficiency'', reports an estimated policy effect corresponding to a reduction of $2.53$pp in child mortality with an interval length reduction of about $3\%$, while the last column reports a reduction of $2.51$pp in child mortality with a slightly larger interval length reduction. All these estimated effects are highly statistically significant and quite similar to each other in magnitude. The models are nested: the first column does not include covariate adjustment, the next column adds eight covariates in the estimation procedure, and then the last column includes one additional covariate for efficiency purposes (\textit{population}). The last column corresponds to the specification used in \cite{Calonico-Cattaneo-Farrell-Titiunik_2019_RESTAT}.

The Head Start application highlights some important empirical and methodological regularities for the role of covariate adjustment for efficiency purposes in RD designs. First, and perhaps most importantly, the point estimates should not change from a statistical (and hence substantive) perspective. Second, the bandwidth employed should change to account for the additional covariates entering the estimation procedure. Third, the average length of the confidence intervals should decrease as more relevant pre-intervention covariates are included in the estimation. These implications are a consequence of properly employing pre-intervention covariates in modern RD estimation and inference. Importantly, it is always good practice to check that the covariates used are indeed unaffected by the policy: this is often done by producing RD plots and conventional RD estimates using the pre-intervention covariate as an outcome. For example, see our companion replication files for the analog of Figure \ref{fig:RDplot-conventional} and the first column in Table \ref{tab:cov-efficiency} using \textit{population} as the outcome variable.

A limitation of leveraging covariates for efficiency is related to p-hacking, that is, an adversarial researcher could exploit the presence of many pre-intervention covariates to isolate the one local least squares regression specification that delivers statistically significant results, disregarding the fact that many regression models were fitted during the search process. This problem is not specific to RD designs \citep[see][]{Blanco-Perez2019-IZAWOL}, and is fortunately mitigated by noting that the point estimates obtained from the canonical and the covariate-adjusted models should be similar to begin with. Thus, researchers should always report both estimates, canonical and covariate-adjusted, together with their accompanying robust bias-corrected statistical inference results. See Table \ref{tab:cov-efficiency} for one example.

\subsection*{Leveraging Covariates for Heterogeneity}

The second main role of covariates in the RD design is for heterogeneity analysis. In this case, the target causal parameter of interest changes, and the specific covariate used plays a different substantive role: the goal is to learn an RD treatment effect for each value of the covariates. Crucially for identification, the covariate used for this type of analysis also needs to be pre-determined, requiring the treatment not to affect it. Estimation and inference methods are different from those discussed in the preceding section, which has led to the proliferation of applied approaches to leverage covariates for heterogeneity analysis in RD designs. Due to the curse of dimensionality, it is often the case that empirical work focuses on heterogeneity analysis, one covariate at a time. Thus, the two most common empirical approaches are (i) to subset the data according to a discrete covariate, or otherwise discretized continuous covariate, and (ii) to incorporate a continuous covariate via joint local linear regression estimation with interactions between the treatment assignment variable and the covariate.

Perhaps surprisingly, there is only a handful of papers studying the methodological aspects of covariate adjustment for heterogeneity analysis in RD designs (see further reading for references). \cite{Calonico-Cattaneo-Farrell-Palomba-Titiunik_2025_wp} is the first paper to study the most common empirical practice based on estimation and inference via local least squares regressions with interactions. Their results are substantively different than, and therefore complement rather than supplant, those in \cite{Calonico-Cattaneo-Farrell-Titiunik_2019_RESTAT}.

We illustrate how to incorporate a covariate to conduct valid heterogeneity analysis using the Head Start data, following the methods recommended by \cite{Calonico-Cattaneo-Farrell-Palomba-Titiunik_2025_wp}. To simplify the discussion, we only study heterogeneous RD treatment effects for small and large counties as determined by their population size: we discretize \textit{population} into a binary variable indicating whether a county has at least 10k residents (henceforth, defined ``large" counties), and then estimate the (conditional) RD treatment effects for each of the two subgroups with and without covariate adjustments for efficiency using the other eight pre-intervention covariates. Before starting with the empirical analysis, we test whether the probability of being a large county is continuous at the cutoff, an important empirical falsification test. We estimate the RD effect using the indicator for large counties as the outcome and find an estimated effect of $0.035$ with a $p$-value of $0.578$. The replication files provide details and generate the corresponding RD Plot. 

\begin{figure}[!ht]
    \caption{Head Start Heterogeneous RD Average Treatment Effects (\textit{population})\label{fig:RDplot-HTE}}
    \centering
    \subfigure[Heterogeneous RD (global)]{
        \includegraphics[width=0.45\textwidth]{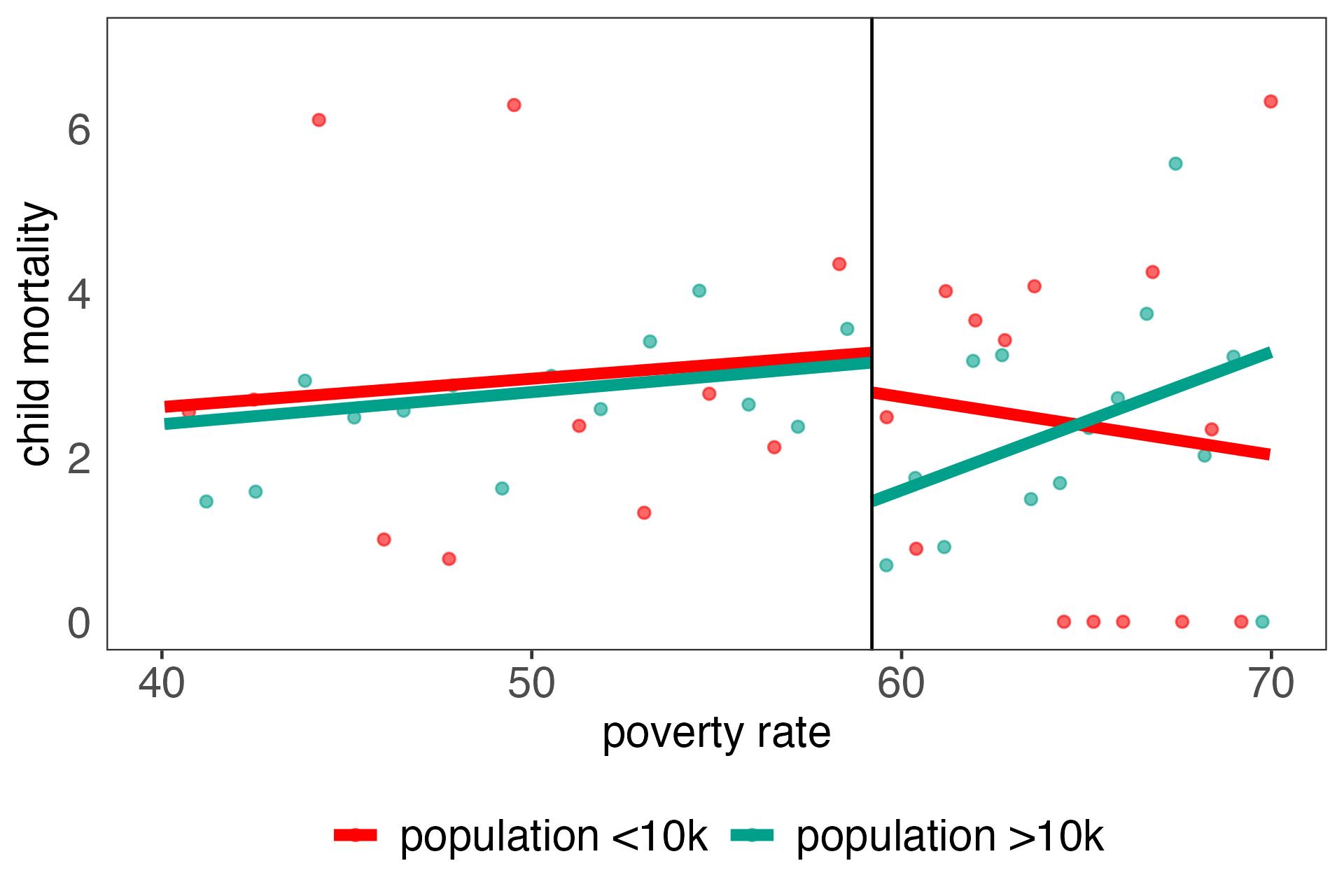}
    } 
    \subfigure[Heterogeneous RD (local)]{
        \includegraphics[width=0.45\textwidth]{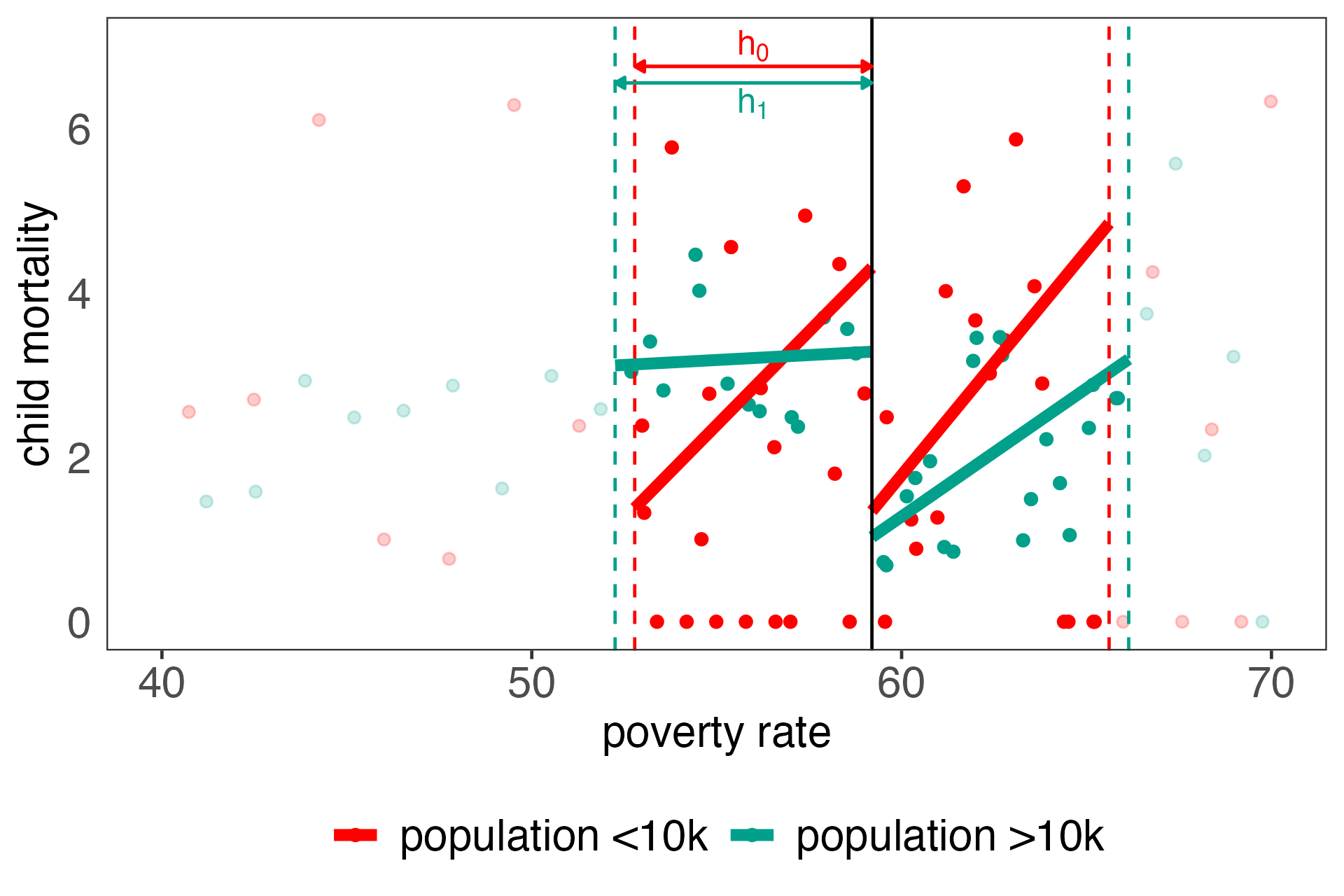}
    }
    \par
    \begin{center}
        \parbox[1]{\textwidth}{\footnotesize \textit{Notes:} In all cases, we use a triangular kernel, use a polynomial of degree 1 in the running variable, and equate the main bandwidth and the auxiliary bias bandwidth. No covariates are included for efficiency. Dots represent averages within equally-spaced bins, solid lines represent separate linear fits on each side of the cutoff, and vertical dashed lines (when displayed) delimit the estimation sample of length twice the optimal MSE bandwidth. Sample sizes for small counties (\textit{population} $<10$k) in panel (a) are $n_-=685$ and $n_+=101$, whereas in panel (b) $N_{-}=71$ and $N_{+}=57$. Sample sizes for large counties (\textit{population} $\geq10$k) in panel (a) are $n_-=1819$ and $n_+=199$, whereas in panel (b) $N_{-}=164$ and $N_{+}=128$.}
    \end{center} 
\end{figure}

To begin the heterogeneity analysis, Figure \ref{fig:RDplot-HTE}(a) presents a global RD Plot for the two subgroups with linear regressions. The plot suggests a more sizable effect for the large counties when compared to the small counties. The first two columns in Table \ref{tab:hte} correspond to the heterogeneous RD analysis when no additional covariates are added for efficiency reasons. The two panels in Table \ref{tab:hte} showcase the results when the bandwidth for small and large counties differ (Panel A) and when they are imposed to be equal (Panel B). Allowing for different bandwidths, as in Panel A, can be useful when the sub-groups are of different sample sizes, whereas imposing the same bandwidth (Panel B) is sometimes desirable for comparability across groups. Nevertheless, the latter approach is generally discouraged, as the bias-variance trade-off that governs the selection of the optimal bandwidth may differ across subsamples of the data, much as it may vary across distinct datasets.

The point estimates for the two subgroups are similar, and the RD treatment effect is statistically significant only for large counties. Specifically, it is found that Head Start decreases child mortality among highly populated counties by roughly $70\%$ relative to the average mortality rate of highly populated control counties. A formal robust $F$-test fails to reject the null hypothesis that the RD treatment effect differs for small and large counties at the 5\% level. Figure \ref{fig:RDplot-HTE}(b) gives a local RD Plot depicting these empirical findings. 

The last two columns in Table \ref{tab:hte} present a heterogeneous RD analysis leveraging the other pre-intervention covariates for efficiency. In this specific illustration, adding the other available covariates to improve precision does not help much. This is an empirical example where the additional covariates are likely to be uncorrelated with the treatment within each subgroup, thereby not offering noticeable efficiency improvements. Furthermore, in this application, studying heterogeneous RD treatment effects leads to relatively small sample sizes, which calls for extra care in the analysis and interpretation of the results.

The Head Start application showcases one instance of heterogeneity analysis in RD designs, where the findings indicate statistically significant policy effects for a particular subpopulation (largely populated counties). It also demonstrates an important limitation often encountered when estimating heterogeneous treatment effects: sample sizes are reduced because subpopulations are considered, hence leading to less precise policy estimates.

\subsection*{Leveraging Covariates for Other Purposes}

Going beyond efficiency improvements and heterogeneity analysis, covariates have been used for other goals in RD designs. We overview these goals; see \cite{Cattaneo-Titiunik_2022_ARE} for references. These methods cannot be illustrated using the Head Start application because they require a specific data structure or other departures beyond the canonical RD design setup to be implemented.

Pre-intervention covariates can enhance RD designs by addressing issues like missing data, measurement error, and incorporating prior information via Bayesian methods. Measurement error in the RD running variable is a common concern, and recent approaches use covariates for correction. Under the assumption of missing-at-random, imputation methods can be applied directly within the local randomization framework. Bayesian methods also offer a distinct approach, leveraging covariates to inform priors or model distributions flexibly. For instance, prior experimental data have been used to inform Bayesian priors in RD studies, while principal stratification has been applied to estimate certain average treatment effects. Although not yet widely adopted, these methods illustrate diverse roles for covariate adjustment in RD applications, primarily to refine estimation without altering the treatment effect parameter of interest.

Covariate adjustments in RD designs are also used to identify additional parameters of interest, such as those for certain subpopulations or for extrapolating treatment effects away from the cutoff. For example, in some RD designs, covariates define new subpopulations through varying cutoffs, as in Multi-Cutoff RD designs. This approach explores heterogeneity by analyzing RD effects separately for each subgroup defined by its cutoff, which can also be aggregated through normalization and pooling under additional assumptions. These ideas are somewhat similar, but often conceptually distinct, from those related to heterogeneity analysis based on pre-intervention covariates as discussed previously. Similarly, treatment dosage changes across multiple cutoffs have been used to explore counterfactual policies and to define effects for subpopulations exposed to different cutoffs, both facilitating RD treatment effect extrapolation. Furthermore, in RD designs spanning multiple periods, treatment confounding can occur in one period but not another. The ``differences in discontinuities'' design estimates effects by comparing RD effects across periods, leveraging assumptions akin to parallel trends for unobservables across time. More recently, in dynamic RD designs, time plays an important role as the key additional covariate determining different treatment regimes.

When multiple cutoffs or periods are unavailable, extrapolation can still rely on auxiliary covariates. A local conditional independence assumption has been used for extrapolation, while in geographic RD designs, assumptions focus on neighborhoods near geographic boundaries. For instance, strategic location choices have been leveraged after imposing conditional independence within boundary neighborhoods (a local rather than conditional independence assumption). Covariates also enhance extrapolation in other ways: RD designs have been augmented with exogenous outcome measures to estimate treatment effects beyond the cutoff under suitable assumptions about outcome regression functions. See further reading for background references.

The ideas reviewed so far showcase a rich literature demonstrating how covariates in RD designs expand the scope of analysis by defining new parameters for subpopulations or treatment effects. These covariates may be inherent to the RD setup, such as multiple cutoffs or periods, or external, requiring separate collection, such as auxiliary characteristics or outcomes. Regardless, they can often be leveraged to improve scientific discoveries and evidence-based policy decisions.

Finally, it is sometimes claimed in the RD literature that covariates can be leveraged to ``fix'' a ``broken'' design. The rationale for adjusting covariate imbalances in RD designs draws on an analogy with flawed randomized experiments, settings where noticeable disparities in pre-intervention covariates arise, either due to defective randomization or random chance. This observation has prompted some researchers to employ covariate adjustment techniques to address shortcomings in the identification of canonical RD treatment effects. However, RD designs present an important added complexity: a lack of common support, which necessarily requires non-trivial extrapolation. In the empirical literature, RD analyses sometimes include fixed effects or other covariate adjustments, typically implemented as linearly additive regression terms in a (local) polynomial regression framework. Methodological advancements have proposed alternative approaches to manage covariate imbalances, such as multi-step non-parametric regression and inverse probability weighting. These techniques, developed within the continuity-based RD framework, aim to address cases where the conditional expectations of covariates given the running variable exhibit discontinuities near the cutoff, reflecting imbalances. As a result, they necessarily change the estimand.

\section*{Limitations and gaps}

The RD design is a well-established methodology for program evaluation and causal inference, supported by numerous falsification tests and robustness checks to assess its validity \cite{Abadie-Cattaneo_2018_ARE,Cattaneo-Titiunik_2022_ARE}. However, several important caveats should be noted. First, the policy effects identified by the RD design are ``local" in nature, meaning they pertain only to units near the cutoff, rather than the entire population. Second, when pre-intervention covariates are included in an RD design, careful interpretation is required, and researchers must clearly articulate their intentions regarding these covariates. Third, examining heterogeneity in policy effects may significantly reduce the sample size available for analysis, which can affect statistical power. Finally, covariate adjustment cannot ``fix" an RD design without imposing strong assumptions and, often, changing the estimand.

\section*{Summary and policy advice}

We examined the different roles that pre-treatment covariates can play in RD designs: enhancing the precision of estimates, exploring heterogeneity in treatment effects, and more broadly changing the estimand. We provided a detailed discussion of each of these roles and illustrated the first two by revisiting a classic empirical study in the RD literature. Additionally, we emphasized the importance of clearly stating the purpose for including pre-treatment covariates in the analysis, as this clarifies the researcher's goal and facilitates the interpretation of empirical findings. Finally, we recommend always conducting empirical falsification testing to enhance the credibility of the findings and also relying on well-developed methods for leveraging covariates in the analysis. Relying on the empirical methods reviewed in this article will help replicability and comparability across studies.

\section*{Acknowledgments}

We thank Sebastian Calonico, Max Farrell, Luke Keele, Rocio Titiunik, and Jeff Wooldridge for inspiring discussions. Cattaneo gratefully acknowledges financial support from the National Science Foundation through grant SES-2241575. 

\section*{Competing interests}

The IZA World of Labor project is committed to the \href{https://legacy.iza.org/en/webcontent/about/IZAResearchIntegrity.pdf}{IZA Guiding Principles of Research Integrity}. The authors declare that they have observed these principles.

\clearpage
\section*{Figures}

\clearpage
\section*{Tables}

\renewcommand{\arraystretch}{1.1}
\begin{table}[!ht]
  \centering
  \caption{Head Start Canonical RD Average Treatment Effect}  \label{tab:cov-efficiency}%
    \resizebox{\textwidth}{!}{
    \input{tables/estimatesAll.tex}}
    \par
    \begin{center}
        \parbox[1]{\textwidth}{\footnotesize \textit{Notes:} In all cases, we use a triangular kernel, a polynomial of degree 1 in the running variable, an HC3-type heteroskedasticity-robust estimator, and equate the estimation bandwidth and the bandwidth to estimate the bias. Column 2 contains results for the case in which no additional covariates are added to achieve efficiency gains; Column 3 shows the same statistics when 8 covariates are added (population between 14 and 17, population between 5 and 34, population with age 25+, \% attending school between 14 and 17, \% attending school between 5 and 34, \% of urban areas, \% of Black population); Column 4 also adds population as a covariate for efficiency gains. The rows report in order: the RD estimator; the bias-corrected 95\% robust confidence interval; the percentage reduction in the RCI compared to the case without covariates; the $p$-value for the null that the true coefficient is 0; the optimal bandwidth; and the effective sample size to the left and right of the cutoff.}
    \end{center} 
\end{table}%

\begin{table}[!ht]
  \centering
  \caption{Head Start Heterogeneous RD Average Treatment Effects}  \label{tab:hte}
    \resizebox{\textwidth}{!}{
    \input{tables/estimatesHte.tex}}
    \par
    \begin{center}
        \parbox[1]{\textwidth}{\footnotesize \textit{Notes:} Panel A reports the results of the heterogeneity analysis obtained using two separate bandwidths for small and large counties. Panel B shows similar statistics using a single bandwidth. Columns 2 and 3 contain results for the case where no additional covariates are added to achieve efficiency gains, whereas Columns 4 and 5 show the same statistics when 8 covariates are added (population between 14 and 17, population between 5 and 34, population with age 25+, \% attending school between 14 and 17, \% attending school between 5 and 34, \% of urban areas, \% of Black population). Within each panel, the table shows in order: the RD estimator; the bias-corrected 95\% robust confidence interval; the percentage reduction in the RCI compared to the case without covariates; the $p$-value for the null that the true coefficient is 0; the optimal bandwidth; the effective sample size to the left and right of the cutoff; and the $p$-value for the test for the null that the RD treatment effect is identical in areas with population above and below the 10k threshold. In all cases, we use a triangular kernel, a polynomial of degree 1 in the running variable, an HC3-type heteroskedasticity-robust estimator, and equate the estimation bandwidth and the bandwidth to estimate the bias.}
    \end{center}  
\end{table}%

\clearpage
\bibliography{CP_2025_IZAWOL--bib}
\bibliographystyle{plain}

\end{document}

%% file: tables/estimatesAll.tex
\begin{tabular}{lccc}
\toprule\toprule
type of RD     & canonical & with covariates for efficiency & with all covariates \\
\midrule
$\hat{\tau}$ & $-2.43$ & $-2.53$ & $-2.48$ \\
95\% RCI & $[-6.33,-1.22]$ & $[-6.60,-1.56]$ & $[-6.48,-1.45]$ \\
CI length change (\%) & -     & $-1.37$ & $-1.63$ \\
$p$-value & 0.004 & 0.002 & 0.002 \\
$h$   & 6.717 & 6.765 & 6.973 \\
$N_-\,|\, N_+$ & $231 \,|\, 179$ & $231 \,|\, 179$ & $240 \,|\, 184$ \\
\% treatment effect & $-78.91$ & $-82.20$ & $-81.70$ \\
\bottomrule\bottomrule
\end{tabular}%

%% file: tables/estimatesHte.tex
\begin{tabular}{lccccc}
\toprule\toprule
  type of RD     & \multicolumn{2}{c}{canonical} &       & \multicolumn{2}{c}{with covariates for efficiency} \\
  \cmidrule{2-3}\cmidrule{5-6} 
      & population $<$10k & population $>$10k &       & population $<$10k & population $>$10k \\
\midrule
\textit{Panel A: different bandwidths} & & &       &       &   \\
\cmidrule{1-1}
$\widehat{\tau}(z)$ & $-2.96$ & $-2.25$ &       & $-4.56$ & $-2.48$ \\
95\% RCI & $[-11.64,2.07]$ & $[-5.96,-0.84]$ &       & $[-11.36,4.59]$ & $[-6.85,-1.31]$ \\
CI length change (\%) & -     & -     &       & $16.39$ & $8.22$ \\
$p$-value & 0.171 & 0.009 &       & 0.406 & 0.004 \\
$h$   & 6.413 & 6.943 &       & 5.004 & 6.304 \\
$N_-\,|\, N_+$ & $71\, | \,55$ & $164\, | \,128$ &       & $56\, | \,45$ & $145\, | \,116$ \\
\% treatment effect & $-103.59$ & $-70.50$ &       & $-157.96$ & $-140.51$ \\
$p$-value (heterog). & \multicolumn{2}{c}{$0.710$}       &       & \multicolumn{2}{c}{$0.871$}  \\
      &       &       &       &       &   \\
\textit{Panel B: same bandwidths} & & &  &       &   \\
\cmidrule{1-1}
$\widehat{\tau}(z)$ & $-2.76$ & $-2.28$ &       & $-3.11$ & $-2.39$ \\
95\% RCI & $[-11.98,1.94]$ & $[-6.13,-0.91]$ &       & $[-14,1.93]$ & $[-6.48,-1.16]$ \\
CI length change (\%) & -     & -     &       & $14.43$ & $2.00$ \\
$p$-value & 0.158 & 0.008 &       & 0.138 & 0.005 \\
$h$   & 6.717 & 6.717 &       & 6.765 & 6.765 \\
$N_-\,|\, N_+$ & $74\, | \,56$ & $157\, | \,123$ &       & $74\, | \,56$ & $157\, | \,123$ \\
\% treatment effect & $-100.72$ & $-70.25$ &       & $-113.48$ & $-95.93$ \\
$p$-value (heterog). & \multicolumn{2}{c}{$0.694$}       &       & \multicolumn{2}{c}{$0.605$}  \\
\bottomrule\bottomrule
\end{tabular}%